\documentclass[fleqn,twoside,twocolumn,nofootinbib]{revtex4} % Specifies the document class %,unsortedaddress
\usepackage[sec]{ujp} % \usepackage[cyr]{ujp} for cyrillic, \usepackage[web]{ujp} for web
\newcommand{\A}{\sigma}

\begin{document}
\title[DIRECT TWO-PHOTON EXCITATION OF ISOMERIC TRANSITION]%колонтитул
{DIRECT TWO-PHOTON EXCITATION OF ISOMERIC TRANSITION IN THORIUM-229 NUCLEUS}%
\author{V.I. ROMANENKO}%1 автор
\affiliation{Institute of Physics, Nat. Acad. of Sci. of
Ukraine}%институт
\address{46, Nauky Ave., Kyiv 03680, Ukraine}%адрес
\email{vr@iop.kiev.ua}%e-mail
\author{Ye.G. UDOVITSKAYA}%1 автор
\affiliation{Institute of Physics, Nat. Acad. of Sci. of
Ukraine}%институт
\address{46, Nauky Ave., Kyiv 03680, Ukraine}%адрес
\email{vr@iop.kiev.ua}%e-mail
\author{L.P. YATSENKO}
\affiliation{Institute of Physics, Nat. Acad. of Sci. of
Ukraine}%
\address{46, Nauky Ave., Kyiv 03680, Ukraine}%
\email{vr@iop.kiev.ua}
\author{A.V.~ROMANENKO}%1 автор
\affiliation{Taras Shevchenko National University of Kyiv}%институт
\address{4, Prosp. Academician Glushkov, Kyiv 03022, Ukraine}%адрес
\email{alexrm@univ.kiev.ua}%e-mail
\author{A.N. LITVINOV}%1 автор
\affiliation{St. Petersburg State Polytechnical University}%институт
\address{29, Politekhnicheskaya Str., St. Petersburg 195251, Russia}%адрес
\email{andrey.litvinov@mail.ru}%e-mail
\author{G.A. KAZAKOV}%1 автор
\affiliation{Institute of Atomic and Subatomic Physics, Vienna
University of
Technology}%институт
\address{Vienna 1020, Austria}%адрес
\email{kazakov@thorium.at}%e-mail
\affiliation{St. Petersburg State Polytechnical University}%институт
\address{29, Politekhnicheskaya Str., St. Petersburg 195251, Russia}%адрес

 \udk{535.372} \pacs{32.50.+d, 32.80.Rm}

%\makeatletter
%\renewcommand{\thesection}{\arabic{section}}
%\renewcommand{\p@subsection}{}
%\renewcommand{\thesubsection}{\arabic{section}.\arabic{subsection}}
%\renewcommand{\p@subsubsection}{}
%\renewcommand{\thesubsubsection}
%{\arabic{section}.\arabic{subsection}.\arabic{subsubsection}}
%\makeatother

%\input{tcilatex}

\razd{\seciv}
\setcounter{page}{1119}%
\maketitle

\begin{abstract}
A possibility of the two-photon excitation of an isomeric state in a
nucleus of thorium-229 has been discussed. The fluorescence
intensity of the excitation is demonstrated to be identical for the
irradiation of nuclei with either monochromatic light or
polychromatic radiation consisting of a sequence of short light
pulses of the same intensity. The two-photon excitation of
$\mathrm{Th^{3+}}$ ion in an electromagnetic trap with a focused
laser beam with a wavelength of about $320$~nm and  power of 100~mW
can lead to the absorption saturation, at which the fluorescence
emission with the frequency of the transition in a nucleus is
maximal. In crystals doped with $\mathrm{Th}^{4+}$ to a
concentration of about $10^{18}$~cm$^{-3}$ and irradiated with a
laser radiation 10~W in power, the emission of several photons per
second with a wavelength of about $160$~nm becomes possible.
\end{abstract}

\section{Introduction}

\label{introduction}A permanent interest to works dealing with the
creation of quantum-mechanical frequency standards (atomic clocks)
is stimulated by both the development of fundamental science and
engineering demands. Modern methods used for the determination of
time unit, a second, are associated with the cesium atomic frequency
standard. In particular, one second is defined as the duration of
9\,192\,631\,770~periods of the radiation corresponding to the
transition between the two hyperfine levels of the ground state of a
$^{133}$Cs atom~\cite{Mar58}. The accuracy of the primary cesium
frequency standard based on the cesium fountain clock was
$4\times10^{-16}$~\cite{Par10}. A possibility for atoms and ions to
be cooled down to ultralow temperatures allowed a stability of the
order of $10^{-17}$ to be obtained using cold Al$^{+}$
ions~\cite{Cho10}. Atomic standards are of interest for navigating
systems, such as GPS, GLONASS, and GALILEO, as well as for
telecommunication networks.

A promising candidate for the role of reference quantum-mechanical frequency
standard is the isomeric transition in a $^{229}$Th nucleus. This isotope
has an extremely low energy of the isomeric state. According to the most recent
data, it equals $7.8\pm 0.6~\mathrm{eV}$ \cite{Bec09}, which corresponds to
a radiation wavelength of about 160\textrm{~nm}. Progress in the development
of the generation of high laser-radiation harmonics allows the nuclear standard
to be regarded as a real possibility. For instance, in work \cite{Pet09},
radiation with a wavelength of 205\textrm{~nm} was obtained with the use of
the generation of the fourth harmonic of radiation emitted by a
titanium-sapphire laser, which is close to that required for the
implementation of a nuclear standard. The development of a nuclear standard is
also stimulated by the capability of its application for studying the
evolution of the ratio between fundamental constants~\cite{Fla06,Lit09}. It
is so because the $^{229}$Th-based frequency standard is supposed to be by
several orders of magnitude more sensitive to a possible variation of
the hyperfine structure constant than the frequency standards based on the
transitions in the electron shell of atoms or ions. Analogously to their
atomic counterparts, nuclear clocks, if being realized, could be applied to
metrology, spectroscopy, global navigating systems, and so forth.

It should be noted that, till now, the energy of the isomeric transition is
known only from the analysis of $\gamma $-spectra with energies that
considerably exceed the nuclear isomeric transition energy. Therefore, it
cannot be considered as ultimately known, until direct measurements of the
level have not been carried out. For this purpose, rather a large number of
thorium nuclei must be excited simultaneously. At the present time, either
thorium ions in traps~\cite{Pei09,Por09,Cam09,Cam11} or thorium-doped
crystals transparent in the ultra-violet range~\cite{Rel10,Kaz11} are
proposed to be used. The advantage of traps consists in the possibility of
a better control over the fields that act on thorium ions. At the same time,
crystals allow a much larger number of nuclei to be
engaged simultaneously than those in traps.

Instead of using the radiation with a frequency close to that of the
transition in a nucleus, the latter can be excited with the use of
two- or multiphoton processes. The choice of that or another
excitation way depends on the parameters of available laser
radiation sources, such as the intensity, spectral width, and
generation mode (pulsed or continuous). Owing to the short
wavelength of radiation needed for the excitation of a nucleus, this
radiation can be obtained by generating the radiation harmonics of
pulsed lasers operating in the visible spectral range. At the
repetition frequency of an order of 100~MHz and under single-photon
excitation conditions, only one of the spectral components, which is
close to the frequency of the transition in a nucleus, can
expectedly stimulate a transition between nuclear states. This means
that, if the radiation emitted by picosecond and femtosecond lasers
and characterized by a considerable number of spectral components is
used as the pumping one, only a small fraction of the emitted
intensity can be used for the nuclear excitation. At the same time,
in the case of two-photon excitation, almost all spectral components
(grouped in pairs consisting of the frequencies locating above and
below the transition frequency and at the same distance from it)
stimulate the nuclear excitation. On the other hand, the probability
of a two-photon transition is much lower than that of a
single-photon one, and a possibility to develop a frequency standard
on the basis of a two-photon transition is not so evident.

It should be noted that the two-photon excitation, in contrast to the single-photon
one, is insensitive, to an accuracy of the squared Doppler effect, to the
atomic velocity, if counter-propagating waves are used, i.e. the excitation by a
standing wave is executed~\cite{Vas70}. Really, if an atom that moves with
the velocity $v$ along the $z$-axis is excited by a standing wave with the
frequency $\omega $, it is subjected to the action of two monochromatic
waves with the frequencies $\omega \pm kv$, where $k=2\pi \omega /c$, in the
accompanying reference frame. Absorption of two photons from the
counter-propagating waves results in an increase of the atomic energy by the
quantity $\hbar (\omega +kv)+\hbar (\omega -kv)=2\hbar \omega $, which is
independent of the atomic velocity.

An interesting method of isomeric transition excitation with the use of two
photons was proposed in work \cite{Por09}. The cited authors suggested to
populate an intermediate level in the electron shell of Th$^{+}$ ion, the
energy of which is well-known, with the help of a laser with narrow
spectral width. The frequency of the second laser is scanned near the
difference between the predicted frequency of the isomeric transition and the
frequency of the first laser. If this procedure gives rise to the excitation
of a level in the ionic electron shell with the energy close to that of
the isomeric transition, this energy can be transferred to the nucleus. In this
case, owing to the variation in the nucleus state, the position of the ionic
intermediate level also changes, so that the first laser field ceases to
interact with it, and the fluorescence at the first-laser frequency diminishes.
At the same time, the issue concerning the lifetime of the isomeric state in the
single-charged ion is not clear in detail \cite{Por10}. Note that the
resonance width for such an excitation is determined by the lifetime of
the intermediate level. Therefore, the method of two-photon nuclear excitation
proposed in work~\cite{Por09} does not give any advantages while developing
a frequency standard, in which a long-term isomeric state of the nucleus is
excited through the excitation of the electron shell, in comparison with that,
where the electron transition in an atom or ion is excited. However, the
method of excitation of a nucleus into the isomeric state, which was proposed in
work~\cite{Por09}, can undoubtedly facilitate the registration of the optical
nuclear transition.

This work is devoted to the analysis of the \textquotedblleft
direct\textquotedblright\ two-photon nuclear transition in a thorium
nucleus making no allowance for any intermediate levels. Such an excitation
can be carried out, generally speaking, in any thorium ion. We have
considered the excitation of nuclei with either the monochromatic or
polychromatic field created by a light wave or counter-propagating light
waves. In Section~\ref{sec:Th}, the major spectroscopic characteristics of
thorium are described, which are relevant to the transition from the ground
nuclear state into the excited one. Section \ref{sec:Model} contains basic
equations. In Section~\ref{sec:Hamiltonian}, we will derive the effective Hamiltonian,
which forms a basis for the description of the time evolution of slowly
changing, in comparison with the laser radiation period, components of
the probability amplitudes for a state of the nucleus. The two-photon absorption in a
monochromatic field is discussed in Section~\ref{sec:twomono}, and that in
the field created by a sequence of propagating light pulses in Section
\ref{sec:pulses}. In Section~\ref{sec:pulses-c}, the two-photon excitation in the
field of sequences of counter-propagating light pulses is considered. In
Section \ref{sec:Discuss}, the efficiency of the two-photon excitation is
evaluated and discussed. In Conclusions, the results obtained in this
work are summarized.

\section{\boldmath$^{229}$Th
Isomer}

\label{sec:Th}$^{229}$Th isotope is $\alpha $-active with a
half-life period of 7340~year~\cite{Audi03}. In Figure, the available
spectroscopic information concerning the magnetic dipole ($M1$)
transition between the first excited level $I_{e}=3/2$ and the
ground state level $I_{g}=5/2$ in a $^{229}$Th nucleus is
summarized.

Each of two levels with the total momentum $I_{i}$ ($I_{g}=5/2$ and
$I_{e}=3/2$) splits into $2I_{i}+1$ sublevels described by the magnetic
quantum number $m=-I_{i},-I_{i}+1,\ldots {}I_{i}-1,I_{i}$. If the nucleus is
subjected to the action of a constant magnetic field, those levels have
different energies. In a solid, the nucleus undergoes the influence of the
electric field created by atoms of the environment. As a result, there
arises a quadrupole shift of the nucleus energy, which depends on $m$. Let the
nucleus be subjected to the action of linearly polarized laser radiation,
which can depend on the time; in particular, it can be a sequence of pulses
with period $T$. Provided this assumption, it is sufficient to analyze
the two-level model of interaction between the nucleus and the field to
evaluate the efficiency of two-photon excitation.

\begin{figure}[t]% figure* for wide figure, [h] [!] to change the placement
\includegraphics[width=\column]{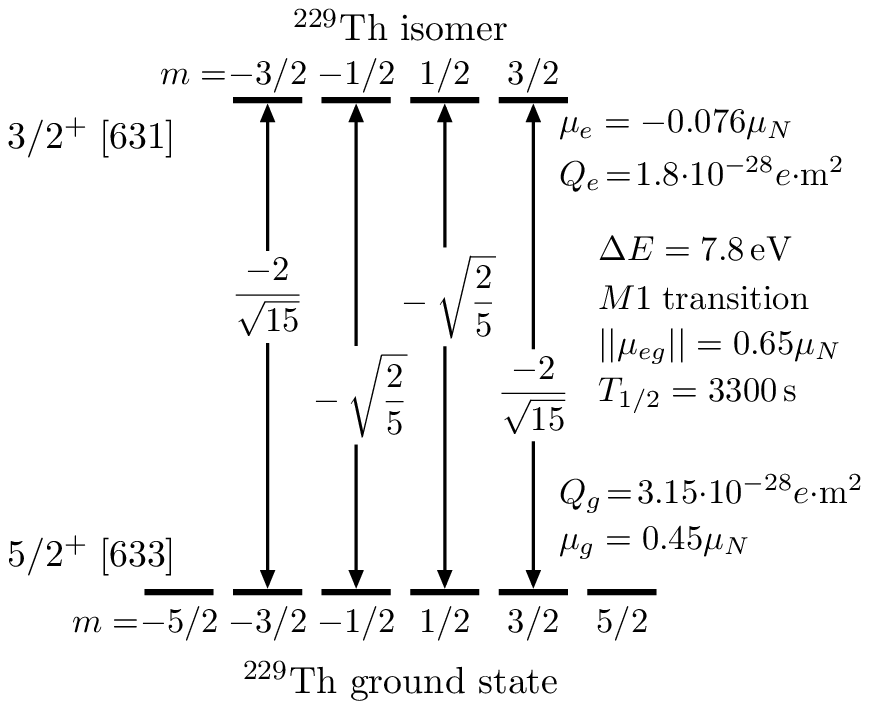}
%В картинке надо поменять данные на более новые!!!
\vskip-3mm\caption{Ground and first excited states of $^{229}$Th
nucleus. The levels are classified according to the Nilsson
model~\cite{Nil55}. The radiation lifetime of a free nucleus
$T_{1/2}=55$~min (the rate of spontaneous radiation emission
$\gamma=0.00021$~s$^{-1}$) and the reduced matrix element of
the transition $\left\Vert \mu_{eg}\right\Vert
=\sqrt{\frac{4\pi}{3}B(M1)(2I_{e}+1)}=0.65\mu_{N}$, where $B(M1)$ is
the reduced probability of decay, were calculated on the basis of
data taken from work \cite{Ruc06}. The magnetic moments of levels in
terms of nuclear magneton units and the quadrupole moment in the
ground state, $Q_{g}=3.15\times10^{-28}~e\cdot\mathrm{m}^{2}$, were
taken from works \cite{Bem98, Dyk98-233}. The estimate
$Q_{e}=1.8\times10^{-28}~e\cdot\mathrm{m}^{2}$ for the quadrupole
momentum in the excited state was taken from work \cite{Tka11}. The
arrows indicate probable transitions between magnetic sublevels in
the field of linearly polarized laser radiation. The Clebsch--Gordan
coefficients, which are proportional to the matrix elements of
the magnetic dipole moments of corresponding transitions, are also
indicated   } \label{fig:scheme}
\end{figure}

Among the whole set of transitions between the atomic or ion states
and the nucleus in the ground or excited state, we consider the
transition between the states with the same orbital moment and spin
of electrons in the ground and excited states. The frequency of this
transition, to an accuracy of corrections associated with the
hyperfine level structure, coincides with that of the transition
between the ground and excited state of a nucleus, i.e. the reference
frequency of optical nuclear frequency standard. Although the state
of electrons remains invariable, their presence induces the
variation of the total moment for the system \textquotedblleft
electron shell + nucleus\textquotedblright, as well as the $g$-factor
responsible for the Stark shift of energy levels in the magnetic
field. A fourfold-charged Th$^{4+}$ ion has a closed electron
shell of radon. Figure illustrates the interaction of this ion with
the field. For Th$^{3+}$ ion, the ground state is $5F_{5/2}$
\cite{Cam11}. As a result, there emerges a hyperfine structure of
the ground state with the total moment $F=0,1,2,3,4,5$ and the
excited state with $F=1,2,3,4$. In a magnetic field, each level of
hyperfine structure splits into $2F+1$ sublevels. Owing to the
Zeeman effect, those sublevels are shifted with respect to their
position in the zero field by
%1
\begin{equation}
\Delta {}E=\mu _{\rm B}g_{F}m_{F}B,
\end{equation}%
where $B$ is the magnetic field induction, and $\mu _{\rm B}$ the
Bohr magneton. The multiplier $g_{F}$ in the case of zero orbital
moment (which is of interest for us) equals
%2
\[
 g_{F}=g_J\frac{F(F+1)-I(I+1)+J(J+1)}{2F(F+1)}+
 \]
\begin{equation}
+g_{I}\frac{F(F+1)+I(I+1)-J(J+1)}{2F(F+1)},
\end{equation}
where $I$ is the nuclear moment, $J$ the angular momentum of a valence
electron, $g_{I}=-\mu _{i}/(\mu _{\rm B}I)$ is the nuclear
$g$-factor, $(i=e,g)$,
%3
\begin{equation}
g_{J}=1+\frac{J(J+1)-L(L+1)+S(S+1)}{2J(J+1)}(g_{e}-1)
\end{equation}%
is the electron $g$-factor, $L$ the orbital moment of electrons, $S$ the
spin, and $g_{e}=2.0023$ is the $g$-factor of a free electron (in the further
calculations, we adopt $g_{e}=2$).

For Th$^{4+}$ ion, $J=0$ and $g_{F}=g_{I}$. For Th$^{3+}$
ion in the ground state, $g_{F}=\frac{3}{7}+\frac{1}{2}g_{I}$ for
every hyperfine sublevel. For Th$^{3+}$ ion with the excited
nucleus ($I=\frac{3}{2}$), similar calculations give, e.g.,
$g_{F}=\frac{15}{28}+\frac{3}{8}g_{I}$ for the hyperfine sublevel
with $F=4$. Since $g_{I}\sim {\mu _{N}}/{\mu _{\rm B}}\ll 1$, the
last term can be neglected. Hence, the $g$-factor for Th$^{3+} $
is by three orders of magnitude larger than that for Th$^{4+}$. As
one can see from what follows, this circumstance brings about a
substantial increase in the efficiency of two-photon excitation of
the thorium nucleus in Th$^{3+}$ ion in comparison with
Th$^{4+}$ one.

\section{Model of Nucleus Interaction with the Field}

\label{sec:Model}In order to estimate the possibility of the direct two-photon
excitation of a thorium nucleus, let us consider transitions between thorium
states characterized by a definite quantum number $m$ related to the ground
and excited nucleus states. If ions are in a solid matrix, their energies
corresponding to different $m$-values are different owing to the interaction
between the ions and the environment. However, if ions or atoms are in the
trap, we consider that a magnetic field eliminating the degeneration with
respect to $m$ is applied to them. Therefore, to analyze the excitation of
a nucleus, we may use the two-level model.

Let the ground state be designated as $\left\vert g\right\rangle $,
and the excited one as $\left\vert e\right\rangle $. The magnetic
dipole transition between the states of an atom in the ground and
excited states has the constant dipole moments $\mu _{gg}=-\mu _{\rm
B}g_{g}m$ and $\mu _{ee}=-\mu _{\rm B}g_{e}m$, where $g_{g}$ and
$g_{e}$ are the $g$-factors for the ground and excited,
respectively, states. The magnetic dipole moment of the transition
between the states $\left\vert g\right\rangle $ and $\left\vert
e\right\rangle $ equals $\mu _{ge}$. Let the atom be subjected to
the action of a field---monochromatic or created by a sequence of
light pulses---with a carrier frequency $\omega $ such that the
value of $2\omega $ is close to the frequency $\omega _{0}$ of
the transition between the states $\left\vert g\right\rangle $ and
$\left\vert e\right\rangle $. Moreover, let the transition in the
nucleus leave the state of atomic electrons intact. In addition, we
suppose that only the terms responsible for the change in the
nucleus state are essential in the Hamiltonian describing the
interaction between the atom and the field.

The induction of the magnetic field of laser radiation with the carrier
frequency $\omega$, which affects the atom, is described by the expression
%4
\begin{equation}
B(t)=\tilde{B}(t)e^{-i\omega{}t}+\tilde{B}(t)^{\ast}e^{i\omega{}t}.
\label{eq:B}
\end{equation}
The probability amplitudes $C_{g}$ and $C_{e}$ to find the nucleus in the
state $\left\vert g\right\rangle $ or $\left\vert e\right\rangle $,
respectively, vary according to the Schr\"{o}dinger equation
%5
\begin{equation}
i\hbar\frac{\partial}{\partial t}{{\boldsymbol{C}}}=H\boldsymbol{C},
\label{eq:Sch}
\end{equation}
where $\boldsymbol{C}$ is the column vector with the components $C_{g}$ and
$C_{e}$, and the Hamiltonian $H$ looks like
%6
\begin{equation}
H=\left( \begin{array}{rr}
-\mu_{gg}B(t) & -\mu_{ge}B(t) \\
-\mu_{eg}B(t) & \hbar\omega_{0}-\mu_{ee}B(t)\end{array}
\right),   \label{eq:H}
\end{equation}
where $\hbar\omega_{0}$ is the energy difference between the states
$\left\vert g\right\rangle $ and $\left\vert e\right\rangle $.

In the general case of a polychromatic field with a narrow, in
comparison with $\omega $, spectral width, expression (\ref{eq:B})
can be interpreted as rapid oscillations with the optical frequency
$\omega $ and a slowly varying amplitude. For a monochromatic field,
$\tilde{B}(t)$ does not depend on the time. For instance, for a
propagating monochromatic wave,
%7
\begin{equation}
\tilde{B}(t)=\tfrac{1}{2}B_{0}e^{ikz},  \label{eq:B-m_i}
\end{equation}%
and, for a monochromatic field of two counter-propagating waves (standing
wave),
%8
\begin{equation}
\tilde{B}(t)=B_{0}\cos {kz},  \label{eq:B-m_ii}
\end{equation}%
where $k=\omega /c$, $z$ is the $z$-coordinate of the atom, and $B_{0}$ is
the wave amplitude.

If the atom is subjected to the action of the field created by a sequence of
light pulses with the pulse-repetition period $T$, the magnetic field
induction at the atom location point can be written down in the form
%9
\begin{equation}
B(t)={B_{0}}\!\!\!\!\sum_{n=-\infty }^{\infty }\!\!\!\!a_{n}\cos \left[
\left( \omega +n\Delta \right) t-k_{n}z+\varphi _{n}\right],  \label{eq:B-p}
\end{equation}%
where $k_{n}=\left( \omega +n\Delta \right) /c$ and $\Delta =2\pi /T$. The
relative amplitudes of spectral components, $a_{n}$, are normalized so that
the component with $n=0$ has the maximum amplitude $a_{0}=1$.

Comparing Eqs.~(\ref{eq:B}) and~(\ref{eq:B-p}), one can see that, for the
field of a sequence of propagating light pulses,
%10
\begin{equation}
\tilde{B}=\tfrac{1}{2}B_{0}\sum_{n=-\infty }^{\infty }a_{n}\exp \left(
-in\Delta {}t+ik_{n}z-i\varphi _{n}\right).  \label{eq:BB-p}
\end{equation}

For the field of the counter-propagating sequences of pulses
%11
\[
 B(t)={B_{0}}\sum_{n=-\infty }^{\infty }a_{n}\cos\left[ \left(
\omega +n\Delta \right) t-k_{n}z+\varphi_{n}\right] +
\]
\begin{equation}
+{B_{0}}\!\!\!\!\sum_{n=-\infty }^{\infty }\!\!\!\!a_{n}\cos\left[
\left( \omega +n\Delta \right) t+k_{n}z+\varphi_{n}\right],
\label{eq:B-p-c}
\end{equation}
we have
%12
\begin{equation}
\tilde{B}=B_{0}\!\!\!\!\sum_{n=-\infty}^{\infty}\!\!\!\!a_{n}\cos{}%
k_{n}z\exp\left( -in\Delta{}t-i\varphi_{n}\right) .
\label{eq:BB-p-c}
\end{equation}
Here, the reference point for the $z$-coordinate is chosen so that,
at the point $z=0$, the field of a wave propagating in the
negative direction of the $z$-axis reproduces the field of the wave
propagating in the positive direction of the $z$-axis.

\section{Effective Hamiltonian}

\label{sec:Hamiltonian}Let us make the substitutions
%13
\[
C_{g}(t)=c_{g}(t)\exp\left(\frac{i\mu_{gg}}{\hbar}\int\limits_{0}^{t}
B(t')\,dt'\right),
\]
\begin{equation}
C_{e}(t)=c_{e}(t)\exp\left(-Ni\omega{}t+\frac{i\mu_{ee}}{\hbar}\int\limits_{0}^
{t} B(t')\,dt'\right). \label{eq:Cc}
\end{equation}
in the Schr\"{o}dinger equation (\ref{eq:Sch}). The variation of
the column vector $\boldsymbol{c}$ with the components $c_{g}$ and
$c_{e}$ in time is described by the Schr\"{o}dinger equation with
the Hamiltonian
%14
\begin{equation}
\mathcal{H}=\left( \begin{array}{cc}
0 & -\mu _{ge}B(t)e^{i\Phi (t)} \\
-\mu _{eg}B(t)e^{-i\Phi (t)} & \hbar \omega _{0}-2\hbar \omega
\end{array}\right),  \label{eq:Hc}
\end{equation}
where
%15
\begin{equation}
\Phi (t)=-2\omega {}t+\frac{\mu _{ee}-\mu _{gg}}{\hbar
}\int\limits^{t}B(t^{\prime })\,dt^{\prime }.  \label{eq:Phi}
\end{equation}%
The lower limit of integration is not indicated, because the proper choice
of the time reference point makes the value of the primitive at this limit equal
to zero.

We consider that the characteristic time $\tau $ of field amplitude
variations (e.g., the pulse duration) considerably exceeds the reciprocal
carrier frequency of the field, so that the inequality
%16
\begin{equation}
\omega \tau \gg 1  \label{eq:tau}
\end{equation}
is valid. Then
%17
\begin{equation} \Phi (t)=-2\omega {}t+\frac{\mu
_{ee}-\mu _{gg}}{i\hbar \omega }\!\!\left[ \tilde{B}(t)^{\ast
}e^{i\omega {}t}-\tilde{B}(t)e^{-i\omega {}t}\right] \!\!.
\label{eq:Phi-p}
\end{equation}%
In particular, for the field of a monochromatic propagating wave ($\tau
=\infty $), we obtain
%18
\begin{equation}
\Phi (t)=-2\omega {}t+B_{0}\frac{\mu _{ee}-\mu _{gg}}{\hbar \omega }\sin
(\omega {}t-kz),  \label{eq:Phi-m-i}
\end{equation}%
whereas, for the monochromatic field of counter-propagating waves,
%19
\begin{equation}
\Phi (t)=-2\omega {}t+2B_{0}\frac{\mu _{ee}-\mu _{gg}}{\hbar \omega }\cos
{kz}\sin \omega {}t.  \label{eq:Phi-m-ii}
\end{equation}

Let us introduce the following functions slowly varying in
time:
%20
\begin{equation}
\beta (t)=\frac{\tilde{B}(t)\mu _{ge}}{\hbar },\qquad \alpha
(t)=\frac{\tilde{B}(t)\left( \mu _{ee}-\mu _{gg}\right) }{\hbar \omega }.
\label{eq:ab}
\end{equation}%
Since $\left\vert \alpha \right\vert \ll 1$, the further calculations will
be carried out to an accuracy of the terms linear in $\alpha $. Substituting
Eqs.~(\ref{eq:Phi-p}) and~(\ref{eq:ab}) into Hamiltonian (\ref{eq:Hc}), we
obtain
%21
\begin{equation}
{\mathcal{H}}=\frac{\hbar }{2}\left(\! \begin{array}{cc}
0 & \Omega (t)+\tilde{\Omega}(t) \\
\Omega (t)^{\ast }+\tilde{\Omega}(t)^{\ast } & 2\delta
\end{array}\!\right).  \label{eq:HH}
\end{equation}%
Here, in the non-diagonal elements, we separated the terms
$\tilde{\Omega}(t)$ rapidly varying in time with the frequency $\omega $ and the slowly
varying terms $\Omega (t)$ and introduced the two-photon detuning
$\delta =\omega _{0}-2\omega $. The fast and slow components of the Rabi
frequency equal
%22
\[
\Omega=-2\alpha^{*}\beta^{*},
\]
\begin{equation}
\tilde{\Omega}=-2\beta^{*}e^{-i\omega{}t}-2\beta{}e^{-3i\omega{}t}.
\label{eg:Rabi-ii}
\end{equation}
Hereafter, in order to simplify the notation and if it cannot lead to
misunderstanding, we do not indicate the time dependence of
quantities. In expression (\ref{eg:Rabi-ii}), we neglected terms
with higher orders of smallness with respect to $\alpha \ll 1$. At
the same time, it should be noticed that only the nonzero value of
$\alpha $ makes the component of the Rabi frequency that slowly varies
in time different from zero for the two-photon process considered
here.

The solution of the Schr\"{o}dinger equation is sought as a sum of two terms,
slowly and rapidly varying in time with a characteristic time of the order
of $2\pi /\omega $, for each of the probability amplitudes,
%23
\begin{equation}
c_{n}(t)=\sigma _{n}(t)+\tilde{\sigma}_{n}(t),\quad {}n=g,e,  \label{eq:sol}
\end{equation}%
where the rapidly varying terms are marked by the tilde sign. We adopt that
the average value of rapidly varying term over the period $2\pi /\omega $
vanishes. From the Schr\"{o}dinger equation, it follows that
%24
\[
i\dot{\A}_{g}+\underline{i\dot{\tilde{\A}}_{g}}={\textstyle\frac{1}{2}}\Omega{}
\A_ {e}
+\underline{\underline{{\textstyle\frac{1}{2}}\tilde{\Omega}\tilde{\A}_{e}}+
{\textstyle\frac{1}{2}}\tilde{\Omega}{\A}_{e}+{\textstyle\frac{1}{2}}{\Omega}
\tilde{\A}_{e}},
\]
\[
i\dot{\A}_{e}+\underline{i\dot{\tilde{\A}}_{e}}={\textstyle\frac{1}{2}}\Omega^{
* }
{}\A_{g}+
\underline{\underline{{\textstyle\frac{1}{2}}\tilde{\Omega}^{*}\tilde{\A}_{g}}+
{\textstyle\frac{1}{2}}\tilde{\Omega}^{*}{\A}_{g}+{\textstyle\frac{1}{2}}{\Omega
} ^{*}\tilde{\A}_{g}}+
\]
\begin{equation}
+\delta{}\A_{e}+\underline{\delta{}\tilde{\A}_{e}}. \label{eq:da}
\end{equation}
The underlined terms rapidly oscillate and vanish after having been averaged
over the oscillation period $2\pi /\omega $. The twice underlined terms
consist of products of oscillating multipliers. They can be presented as a
sum of a component that slowly varies in time and an oscillating component
with the averaged value equal to zero. To exclude the rapid motion, let us
collect together, into two groups, rapidly and slowly oscillating terms in
Eqs.~(\ref{eq:da}). The oscillating terms in Eqs.~(\ref{eq:da}) give
%25
\[
i\dot{\tilde{\A}}_{g}=
{\textstyle\frac{1}{2}}\tilde{\Omega}{\A}_{e},
\]
\begin{equation}
i\dot{\tilde{\A}}_{e}=
{\textstyle\frac{1}{2}}\tilde{\Omega}^{*}{\A}_{g}. \label{eq:daf}
\end{equation}
Here, we took into account that $\left\vert
\dot{\tilde{\sigma}}_{n}(t)\right\vert \sim \omega \left\vert
{\tilde{\sigma}}_{n}(t)\right\vert $, $\omega \gg (\Omega
,\left\vert \tilde{\Omega}(t)\right\vert ,\left\vert \delta
\right\vert )$, and $\left\vert \tilde{\sigma}_{n}(t)\right\vert \ll
1 $. The solution of Eqs.~(\ref{eq:da}) reads
%26
\[
\tilde{\A}_{g}=-\left(\frac{\beta^{*}}{\omega}e^{-i\omega{}t}+\frac{\beta}{
3\omega}e^{-3i\omega{}t}\right)\A_{e},
\]
\begin{equation}
\tilde{\A}_{e}=\left(\frac{\beta}{\omega}e^{i\omega{}t}+\frac{\beta^{*}}{
3\omega }e^{3i\omega{}t}\right)\A_{g}. \label{eq:dafsol-ii}
\end{equation}
The terms in Eqs.~(\ref{eq:da}) that slowly vary in time bring about
%27
\[
i\dot{\A}_{g}={\textstyle\frac{1}{2}}\Omega{}\A_{e}+
\left\langle{\textstyle\frac{1}{2}}\tilde{\Omega}\tilde{\A}_{e}\right\rangle,
\]
\begin{equation}
i\dot{\A}_{e}={\textstyle\frac{1}{2}}\Omega^{*}{}\A_{g}+
\left\langle{\textstyle\frac{1}{2}}\tilde{\Omega}^{*}\tilde{\A}_{g}
\right\rangle+ \delta{}\A_{e}, \label{eq:das}
\end{equation}
where the notation $\left\langle \cdots \right\rangle $ means
the time-averaging over the interval $2\pi /\omega $. After carrying out
this averaging, we obtain
%28
\[
i\dot{\A}_{g}={\textstyle\frac{1}{2}}\Omega{}\A_{e}-\frac{4|\beta|^{2}}{3\omega
} \A_{g},
\]
\begin{equation}
i\dot{\A}_{e}={\textstyle\frac{1}{2}}\Omega^{*}{}\A_{g}+
\frac{4|\beta|^{2}}{3\omega}\A_{e} +\delta{}\A_{e}.
\label{eq:das-av-ii}
\end{equation}
As a result, we find that the variation of the slow components of the probability
amplitudes is described by the Schr\"{o}dinger equation with the effective
Hamiltonian
%29
\begin{equation}
{\mathcal{H}}_{\mathrm{eff}}=\frac{\hbar }{2}\left( \begin{array}{cc}
-S & \Omega  \\ \Omega ^{\ast } & 2\delta
+S\end{array}\right) ,  \label{eq:Heff}
\end{equation}
where
%30
\[
S=
\frac{8|\beta|^{2}}{3\omega}=\frac{8|\tilde{B}\mu_{ge}|^{2}}{3\hbar^{2}\omega},
\]
\begin{equation}
\Omega=-2\alpha^{*}\beta^{*}=-2\frac{{\tilde{B}^{*}}{}^{2}}{\hbar^{2}\omega}
\mu_{ge}(\mu_{ee}-\mu_{gg}). \label{eq:Rabi-shift-ii}
\end{equation}

As one can see from Eq.~(\ref{eq:Rabi-shift-ii}), the light shift is
proportional to the laser radiation intensity. In effect, this relation was
obtained because we have gone beyond the rotating wave
approximation~\cite{Sho90}. The ratio $\left\vert S/\Omega \right\vert \sim
\left\vert \mu _{ge}/(\mu _{ee}-\mu _{gg})\right\vert $ is of the order of
$10^{-3}$ in the case where the $g$-factors of the ground and excited atomic
states are determined by the atomic electron structure, e.g., for a
triple-charged thorium ion in the trap.

\section{Two-Photon Interaction of a Nucleus with a Monochromatic Field}

\label{sec:twomono}To estimate the rate of nucleus fluorescence under the
influence of monochromatic laser radiation, it is convenient to write down
expressions for the Rabi frequency and the light shift in the case where the
nucleus is excited by the propagating monochromatic wave. From
the expression for the radiation intensity in the SI system,
%31
\begin{equation}
I=\frac{c}{2\mu _{0}}B_{0}^{2},  \label{eq:I-i}
\end{equation}%
and Eqs.~(\ref{eq:B-m_i}) and (\ref{eq:Rabi-shift-ii}), we obtain
%32
\[
S=\frac{4|\mu_{ge}|^{2}}{3\hbar^{2}\omega{}c}\mu_{0}I,
\]
\begin{equation}
\Omega=-\frac{\mu_{0}I}{\hbar^{2}\omega{}c}
\mu_{ge}(\mu_{ee}-\mu_{gg})e^{-2ikz}. \label{eq:Rabi-shift-ii-I}
\end{equation}

In the case of monochromatic standing wave, from expressions
(\ref{eq:B-m_ii}), (\ref{eq:Rabi-shift-ii}), and (\ref{eq:I-i}), we
obtain
%33
\[
S=\frac{16|\mu_{ge}|^{2}}{3\hbar^{2}\omega{}c}\mu_{0}I\cos^{2}kz,
\]
\begin{equation}
\Omega=-\frac{4\mu_{0}I}{\hbar^{2}\omega{}c}
\mu_{ge}(\mu_{ee}-\mu_{gg})\cos^{2}kz, \label{eq:Rabi-shift-ii-I-s}
\end{equation}
where $I$ is the intensity of either of the counter-propagating waves.

For an atom or ion moving in the trap, in the case of a propagating
monochromatic wave, the proportionality of the Rabi frequency to the quantity
$e^{-2ikz}$, in which $z$ linearly depends on the time, brings about a
Doppler shift of the resonance by the magnitude of $2kv$, where $v$ is the atom
velocity. This means that, for narrow resonances to be obtained, it is
necessary to cool atoms down to ultralow temperatures. For two
counter-propagating waves (standing light wave), the multiplier
%34
\begin{equation}
\cos ^{2}kz=\frac{1}{2}+\frac{1}{4}e^{2ikz}+\frac{1}{4}e^{-2ikz}
\end{equation}%
in expression~(\ref{eq:Rabi-shift-ii-I-s}) for the Rabi frequency testifies
that there emerge three resonances, when the atom interacts with the field;
namely, two of them are shifted with respect to the transition frequency by
$\pm 2kv$, and one resonance is located at the transition frequency. Since
all groups of atoms make contribution to the latter resonance, the
dependence of the fluorescence intensity on the detuning $\delta $ looks like a
wide peak associated with the Doppler line broadening and a narrow high
peak at its center. This phenomenon serves as a basis for two-photon
spectroscopy~\cite{Vas70,Letokhov}. Therefore, there is no need to use
ultracold atoms or their localization in a small volume (the Dicke effect)
at the two-photon excitation in order to reduce a negative influence of
the Doppler effect on the frequency standard functioning.

\section{Interaction of a Nucleus with the Field of a Sequence of Propagating
Light Pulses}

\label{sec:pulses}Let the atom undergo the action of the field of a sequence
of light pulses with the pulse period $T$. The magnetic field induction at
the point, where the atom is located, is described by expression
(\ref{eq:B-p}). The spectrum of this field is equidistant, with the difference between
the frequencies of spectral components $\Delta =2\pi /T$. On the basis of
expressions (\ref{eq:BB-p}) and (\ref{eq:Rabi-shift-ii}), it is evident that
the light shift and the Rabi frequency can be expanded in the Fourier series
%35
\begin{equation}
S=\sum_{n=-\infty }^{\infty }S_{n}e^{in\Delta {}t}  \label{eq:shift}
\end{equation}
and
%36
\begin{equation}
\Omega =\sum_{n=-\infty }^{\infty }\Omega _{n}e^{in\Delta {}t},
\label{eq:Rabi}
\end{equation}
where
%37
\[
 S_{n}=
\frac{2|{B}_{0}\mu_{ge}|^{2}}{3\hbar^{2}\omega}\sum_{j=-\infty
}^{\infty }a_{j-n}a_{j} \times
\]
\begin{equation}
\times \exp\left(i\varphi_{j}-i\varphi_{j-n}+ik_{j-n}
z-ik_{j}z\right) \label{eq:shiftn-ii}
\end{equation}
and
%38
\[
\Omega_{n}=-\frac{{{B}_{0}}^{2}}{2\hbar^{2}\omega}\mu_{ge}(\mu_{ee}-\mu_{gg})
\sum_{j=-\infty }^{\infty }a_{n-j}a_{j}\times
\]
\begin{equation}
\times\exp\left(i\varphi_{j}+i\varphi_{n-j} -ik_ {n-j}z-ik_{j}z
\right). \label{eq:Rabin-ii}
\end{equation}
The light shift $S$ and the Rabi frequency $\Omega $ include
time-independent terms and terms oscillating with the period $T=2\pi
/\Delta $,
%39
\begin{equation} S=S_{0}+{\tilde{S}},\qquad \Omega =\Omega
_{0}+\tilde{\tilde{\Omega}}. \label{eq:osc}
\end{equation}

We consider the low intensities of laser radiation, for which $\left\vert
S\right\vert \ll \Delta $ and $\left\vert \Omega \right\vert \ll \Delta $.
Then, by applying the procedure of averaging over the rapid motion (in this
case, with the frequency $\Delta $), which was described in
Section~\ref{sec:Hamiltonian}, it is possible to determine the effective Hamiltonian
describing the variation of the slow components of the probability amplitudes
$\sigma _{g}$ and $\sigma _{e}$ in time. Similarly to Eq.~(\ref{eq:sol}), we
write down the quantities $\sigma _{g}$ and $\sigma _{e}$ as sums of slowly
varying in time and oscillating (marked by the tilde sign) terms,
%40
\begin{equation}
\sigma _{j}(t)=b_{j}(t)+\tilde{b}_{j}(t),\quad {}j=g,e.  \label{eq:sola}
\end{equation}%
From the Schr\"{o}dinger equation with Hamiltonian~(\ref{eq:Heff}), it
follows that
%41
\[
i\dot{b}_{g}+\underline{i\dot{\tilde{b}}_{g}}={\textstyle\frac{1}{2}}\Omega{}_{
0}b_{e}
+\underline{\underline{{\textstyle\frac{1}{2}}\tilde{\tilde{\Omega}}\tilde{b}_{e
}}+
{\textstyle\frac{1}{2}}\tilde{\tilde{\Omega}}{b}_{e}+{\textstyle\frac{1}{2}}{
\Omega}_{0}\tilde{b}_{e}}-
\]
\[
-{\textstyle\frac{1}{2}}S_{0}b_{g}
-\underline{\underline{{\textstyle\frac{1}{2}}{\tilde{S}}\tilde{b}_{g}}-
{\textstyle\frac{1}{2}}{\tilde{S}}{b}_{g}-{\textstyle\frac{1}{2}}{S}_{0}\tilde{b
}_{g}},
\]
\[
i\dot{b}_{e}+\underline{i\dot{\tilde{b}}_{e}}={\textstyle\frac{1}{2}}\Omega_{0}
^{*}{}b_{g}+
\underline{\underline{{\textstyle\frac{1}{2}}\tilde{\tilde{\Omega}}^{*}\tilde{b}
_{g}}+
{\textstyle\frac{1}{2}}\tilde{\tilde{\Omega}}^{*}{b}_{g}+{\textstyle\frac{1}{2}}
{\Omega}_{0}^{*}\tilde{b}_{g}}+
\]
\begin{equation}
+{\textstyle\frac{1}{2}}S_{0}b_{e}+
\underline{\underline{{\textstyle\frac{1}{2}}{\tilde{S}}\tilde{b}_{e}}+
{\textstyle\frac{1}{2}}{\tilde{S}}{b}_{e}+{\textstyle\frac{1}{2}}{S}_{0}\tilde{b
}_{e}}+\delta{}b_{e}+\underline{\delta{}\tilde{b}_{e}}.
\label{eq:db}
\end{equation}
Here, analogously to the notation in Eq.~(\ref{eq:da}), the underlined
terms oscillate and vanish after being averaged over the oscillation period,
whereas the twice underlined terms consist of products of oscillating
multipliers and can be presented as sums of slow and rapid components.
Equating the rapid terms independently, we obtain
%42
\[
 i\dot{\tilde{b}}_{g}=
{\textstyle\frac{1}{2}}\tilde{\tilde{\Omega}}{b}_{e}-{\textstyle\frac{1}{2}}
\tilde{S}{b}_{g},
\]
\begin{equation}
i\dot{\tilde{b}}_{e}=
{\textstyle\frac{1}{2}}\tilde{\tilde{\Omega}}^{*}{b}_{g}+{\textstyle\frac{1}{2}}
\tilde{S}{b}_{e}. \label{eq:dbf}
\end{equation}
Here, we took into account that $\left\vert
\dot{\tilde{b}}_{n}(t)\right\vert \sim \Delta \left\vert {\tilde{b}}_{n}(t)\right\vert $,
$\Delta \gg (\Omega _{0},|\tilde{\tilde{\Omega}}|,|\delta |)$, and
$\left\vert {\tilde{b}}_{n}(t)\right\vert \ll 1$. The solution of
Eqs.~(\ref{eq:dbf}) looks like
%43
\[
\tilde{b}_{g}=
\frac{1}{2\Delta}\sum\limits_{\parbox{3em}{\scriptsize$n=-\infty$\newline$n\ne
0$\hfill}}^{\infty}\!\!\!\!
\left({S_{n}b_{g}}-{\Omega_{n}b_{e}}\right)\frac{e^{i\Delta{}nt}}{n},
\]
\begin{equation}
\tilde{b}_{e}=
-\frac{1}{2\Delta}\sum\limits_{\parbox{3em}{\scriptsize$n=-\infty$\newline$n\ne
0$\hfill}}^{\infty}\!\!\!\!
\left({\Omega_{-n}^{*}b_{g}}+{S_{n}b_{e}}\right)\frac{e^{i\Delta{}nt}}{n}.
\label{eq:dbfsol}
\end{equation}
The terms in Eq.~(\ref{eq:db}) that slowly change in time give
%44
\[
i\dot{b}_{g}={\textstyle\frac{1}{2}}\Omega_{0}b_{e}-{\textstyle\frac{1}{2}}S_{0
}b_{g}+
\left\langle{\textstyle\frac{1}{2}}\tilde{\tilde{\Omega}}\tilde{b}_{e}
\right\rangle-
\left\langle{\textstyle\frac{1}{2}}\tilde{S}\tilde{b}_{g}\right\rangle,
\]
\begin{equation}
i\dot{b}_{e}={\textstyle\frac{1}{2}}\Omega^{*}_{0}{}b_{g}+{\textstyle\frac{1}{2
}}S_{0}{}b_{e}+
\left\langle{\textstyle\frac{1}{2}}\tilde{\tilde{\Omega}}^{*}\tilde{b}_{g}
\right\rangle+
\left\langle{\textstyle\frac{1}{2}}{\tilde{S}}\tilde{b}_{e}\right\rangle+\delta{}b_{e},
\label{eq:dbs}
\end{equation}
where the notation $\left\langle \cdots \right\rangle $ means the averaging over
the time interval $2\pi /\Delta $. From expressions (\ref{eq:shiftn-ii}),
(\ref{eq:Rabin-ii}), (\ref{eq:dbfsol}), and~(\ref{eq:dbs}), one can see that,
in the case of two-photon interaction, the averaged terms are quadratic in
the intensity, and they can be neglected. Then the equations for the slowly
varying components of the probability amplitudes are as follows:
%45
\[
i\dot{b}_{g}={\textstyle\frac{1}{2}}\Omega_{0}{}b_{e}-{\textstyle\frac{1}{2}}
\overline{S} b_{g},
\]
\begin{equation}
i\dot{b}_{e}={\textstyle\frac{1}{2}}\Omega_{0}^{*}{}b_{g}+{\textstyle\frac{1}{2
}}\overline{S}b_{e} +\delta{}b_{e}, \label{eq:dbs-av}
\end{equation}
where
%46
\begin{equation}
\overline{S}=S_{0}.  \label{eq:S-ii}
\end{equation}

Hence, the variation in time of the slow components of the probability amplitudes
in the case where the nucleus is excited by a sequence of propagating pulses is described
by the Schr\"{o}dinger equation with the effective Hamiltonian
%47
\begin{equation}
{\mathcal{H}}_{\text{eff}}=\frac{\hbar}{2}\left( \begin{array}{cc}
-\overline{S} & \Omega_{0} \\ \Omega_{0}^{\ast} &
2\delta+\overline{S}\end{array}
\right),   \label{eq:Heffb}
\end{equation}
where $\overline{S}$ and $\Omega_{0}$ are defined by expressions
(\ref{eq:S-ii}) and (\ref{eq:Rabin-ii}) at $n=0$, respectively.

For the further calculations, we must select a model for the field. We assume
that the phases $\varphi _{n}$ of all spectral components of the field equal
zero, and the amplitudes are described by the Gaussian distribution
%48
\begin{equation}
a_{n}=\exp \left( -\frac{n^{2}}{n_{0}^{2}}\right) .  \label{eq:Gauss}
\end{equation}%
Let $n_{0}\gg 1$, so that the summation in the expressions given
above can be replaced by the integration if the time $t$ satisfies
the condition $n\Delta {}t\ll 1$. This inequality means that we do
not deal with time intervals far from light pulses, where the field
strength is very low. The calculation of $\tilde{B}$ gives
%49
\begin{equation}
\tilde{B}=B_{0}\frac{n_{0}\sqrt{\pi }}{2}\exp \left[
ik_{0}z-\frac{n_{0}^{2}\Delta ^{2}}{4}\left( t-\frac{z}{c}\right) ^{2}\right].
\label{eq:B-sum}
\end{equation}%
This expression describes the time dependence of the field created by one of
the pulses that follow one another with period $T$. As one can see from
Eq.~(\ref{eq:B-sum}), the field created by pulses is described by the
Gaussian function $\exp \left( -t^{2}/\tau _{p}^{2}\right) $, where
%50
\begin{equation}
\tau _{p}=\frac{2}{\Delta {}n_{0}}=\frac{T}{\pi {}n_{0}}.  \label{eq:taup}
\end{equation}The time-averaged intensity of laser radiation is equal to
%51
\begin{equation}
I=\frac{2c\langle |\tilde{B}|^{2}\rangle }{\mu _{0}}=\frac{n_{0}c\sqrt{2\pi
^{3}}}{4\mu _{0}}B_{0}^{2}.  \label{eq:I}
\end{equation}
The light shift is determined by the expression
%52
\begin{equation}
\overline{S}=\frac{4\mu _{ge}^{2}\mu _{0}}{3\hbar ^{2}\omega {}c}I.
\label{eq:S0-ii-I}
\end{equation}
The calculation of the Rabi frequency $\Omega _{0}$ gives
%53
\begin{equation}
\Omega _{0}=-\frac{\mu _{ge}(\mu _{ee}-\mu _{gg})}{\hbar ^{2}\omega {}c}\mu
_{0}Ie^{-2ik_{0}z}  \label{eq:Omega0-ii-I}
\end{equation}%
Expression~(\ref{eq:S0-ii-I}) for the light shift and
expression~(\ref{eq:Omega0-ii-I}) for the Rabi frequency coincide with the corresponding
formulas~(\ref{eq:Rabi-shift-ii-I}) obtained for the field of a propagating
monochromatic wave.

The coordinate dependence of the Rabi frequency (\ref{eq:Omega0-ii-I}) for a
moving atom testifies, as was indicated in Section~\ref{sec:twomono} in the
case of propagating monochromatic waves, to the Doppler shift of the resonance
frequency of the interaction of the nucleus with the field by $2kv$. Therefore, in the
case of free atoms or ions, it is necessary to analyze the interaction of
the atom with the field of counter-propagating pulse sequences, for which
the formation of a narrow resonance is possible, at least under the interaction
with a monochromatic field.

\section{Two-Photon Interaction of a Nucleus with the Field of
Counter-Propagating Sequences of Light Pulses}

\label{sec:pulses-c}From expressions (\ref{eq:Rabi-shift-ii}) for the light
shift and the Rabi frequency at the two-photon interaction between the nucleus and
the field of counter-propagating sequences of light pulses (see
Eqs.~(\ref{eq:B-p-c}) and (\ref{eq:BB-p-c})), we obtain
%54
\begin{equation}
S=\sum_{n=-\infty }^{\infty }S_{n}e^{in\Delta {}t}  \label{eq:shift-c}
\end{equation}
and
\begin{equation}
\Omega =\sum_{n=-\infty }^{\infty }\Omega _{n}e^{in\Delta {}t},
\label{eq:Rabi-c}
\end{equation}
where
%56
\[
S_{n}=
\frac{8|{B}_{0}\mu_{ge}|^{2}}{3\hbar^{2}\omega}\!\!\!\!\sum_{j=-\infty
}^{\infty }\!\!\!\!a_{j-n}a_{j} \cos{}k_{j-n}z\times
\]
\begin{equation}
\times\cos{}k_{j}z\exp\left(i\varphi_{j}-i\varphi_{j-n}\right)
\label{eq:shiftn-c}
\end{equation}
and
%57
\[
\Omega_{n}=-2\frac{{{B}_{0}}^{2}}{\hbar^{2}\omega}\mu_{ge}(\mu_{ee}-\mu_{gg})
\!\!\!\!\sum_{j=-\infty }^{\infty }\!\!\!\!a_{n-j}a_{j}\times
\]
\begin{equation}
\times\cos{}k_{n-j}z\cos{}k_{j}z\exp\left(i\varphi_{j}+i\varphi_{n-j}
\right). \label{eq:Rabin-c}
\end{equation}
As was done in Section \ref{sec:pulses}, we normalize the relative
amplitudes of spectral components $a_{n}$ in such a way that the
component with $n=0$ has the maximum amplitude $a_{0}=1$. Carrying
out the calculations similar to those made in Section \ref{sec:pulses},
we determine the effective Hamiltonian (\ref{eq:Heffb}) describing
the two-photon interaction between the atom and the field, where
$\overline{S}=S_{0}$.

Let us choose the same model of field as that in the previous section,
%58
\begin{equation}
a_{n}=\exp \left( -\frac{n^{2}}{n_{0}^{2}}\right) .  \label{eq:Gauss-c}
\end{equation}%
For pico- and femtosecond-pulses, which we are interested in, $n_{0}\gg 1$,
and the summation in the expressions given above can be replaced by the integration
at $n\Delta {}t\ll 1$. The calculation of $\tilde{B}$ gives
%59
\[
\tilde{B}=B_{0}\frac{n_{0}\sqrt{\pi}}{2}\exp\left[ik_{0}z-\frac{n_{0}^{2}
\Delta^{2}}{4}\left(t-\frac{z}{c}\right)^{2}\right]+
\]
\begin{equation}
+B_{0}\frac{n_{0}\sqrt{\pi}}{2}\exp\left[-ik_{0}z-\frac{n_{0}^{2}\Delta^{2}}{4}
\left(t+\frac{z}{c}\right)^{2}\right] \label{eq:B-sum-c}
\end{equation}
for two counter-propagating pulses. The whole sequence of pulses is
obtained from Eq.~(\ref{eq:B-sum-c}) by repeating it with period
$T$. As is evident from Eq.~(\ref{eq:B-sum-c}), the field of pulses
is described by the Gaussian function $\exp \left( -t^{2}/\tau
_{p}^{2}\right) $, where $\tau _{p}$ is determined by expression
(\ref{eq:taup}).

The light shift is calculated from Eq.~(\ref{eq:shiftn-c}) taken at
$n=0$ with regard for Eq.~(\ref{eq:I}) as
%60
\begin{equation}
S_{0}=\frac{8\mu _{ge}^{2}\mu _{0}I}{3\hbar ^{2}\omega {}c}\left(
1+e^{-2z^{2}/l_{p}^{2}}\cos {}2kz\right),  \label{eq:S0-cI}
\end{equation}%
where $l_{p}=c\tau _{p}$, and $I$ is the intensity of either of the
counter-propagating waves. Expression~(\ref{eq:S0-cI}) is valid if
the value of $\Delta z/(2\pi c)$ is close to an integer. Hence, the
light shift consists of two components; one of them is constant in
space, and the other is modulated with the spatial period $\lambda
=2\pi /k_{0}$ and has the envelope in the form of a set of Gaussian
curves located in a vicinity of $z=cTn$, where $n$ is an integer;
here, the counter-propagating pulses \textquotedblleft
collide\textquotedblright .

Similar calculations for the Rabi frequency $\Omega _{0}$ in
Eq.~(\ref{eq:Rabin-c}) give rise to
%61
\begin{equation}
\Omega _{0}=-\frac{2\mu _{0}I}{\hbar ^{2}\omega {}c}\mu _{ge}(\mu _{ee}-\mu
_{gg})\left( \cos {}2kz+e^{-2z^{2}/l_{p}^{2}}\right).  \label{eq:Omega0-cI}
\end{equation}%
The exponential function in Eq.~(\ref{eq:Omega0-cI}) corresponds to the
spatial overlapping of pulses. The component of $\Omega _{0}$ proportional
to $\cos 2kz$ is responsible for the formation of resonances shifted by $\pm
2kv$ owing to the Doppler effect in the case of moving atoms or ions. The
resonance, which is independent of the atom velocity, arises owing to the
term with the Gaussian coordinate dependence. It is connected with a
simultaneous absorption of photons from two counter-propagating pulses. The
$S_{0}$- and $\Omega _{0}$-quantities averaged over the wavelength equal
%62
\begin{equation}
\langle S_{0}\rangle =\frac{8\mu _{ge}^{2}\mu _{0}I}{3\hbar ^{2}\omega {}c}
\label{eq:S0av}
\end{equation}
and
%63
\begin{equation}
\langle \Omega _{0}\rangle =-\frac{2\mu _{0}I}{\hbar ^{2}\omega {}c}\mu
_{ge}(\mu _{ee}-\mu _{gg})e^{-2z^{2}/l_{p}^{2}}.  \label{eq:Omega0av}
\end{equation}%
It is this value of Rabi frequency averaged over the wavelength that is
responsible for the formation of a narrow resonance at the frequency equal to
half a frequency of the transition in the nucleus.

One can see that the magnitude of Rabi frequency averaged over the
wavelength is described by a Gaussian with the maximum located at the point
where the pulses \textquotedblleft collide\textquotedblright. Whence, it
follows that, for the atoms to effectively interact with the field, they must
be localized in a volume with the linear dimension along the direction of
pulse propagation of an order of $l_{p}$. For instance, it can be a cell
filled with a thorium-containing gas, or atoms can be localized in an
optical trap. Two-photon excitation of nuclei with a sequence of laser
pulses is effective as much as that using a monochromatic field, because all
spectral field components are engaged at that. For example, in the resonance
case $2\omega =\omega _{0}$, the pairs $\left( n,-n\right) $ of spectral
components participate in the formation of the two-photon transition.

\section{Discussion of Results}

\label{sec:Discuss}Let us estimate the expected fluorescence of a specimen
in the framework of the two-level model of interaction between the nucleus and
the laser radiation field. Provided that the rate of relaxation of level
populations is equal to $\gamma $ (we assume that the relaxation occurs owing to
the spontaneous radiation emission from the excited state), and their coherence
equals $\gamma ^{\prime }$, the stationary population of the excited state
amounts to
%64
\begin{equation}
\rho _{22}=\frac{1}{2}\frac{|\Omega _{0}|^{2}\gamma ^{\prime }}{|\Omega
_{0}|^{2}\gamma ^{\prime }+\gamma \left( \gamma ^{\prime }{}^{2}+(\delta
+S)^{2}\right) },  \label{eq:sat}
\end{equation}%
where $\delta $ is the detuning from the two-photon resonance. For a free
atom or ion in the field of laser radiation, $\gamma ^{\prime
}=\frac{1}{2}(\gamma +\gamma _{L})$, where $\gamma _{L}$ is the diffusion coefficient for
the laser radiation phase~\cite{Coo80}, i.e. the width of the laser spectrum in
the case of monochromatic radiation. From Eq.~(\ref{eq:sat}), taking into
account that $\gamma \ll \gamma _{L}$, one can see that, for the absorption
to saturate at the resonance, $\delta +S=0$, the condition $\left\vert
\Omega _{0}\right\vert >\sqrt{\gamma \gamma _{L}^{{}}}\sim 0.03$~s$^{-1}$
must be satisfied. For instance, if $\gamma _{L}/2\pi \approx 1$~Hz, it is
obeyed at $\left\vert \Omega _{0}\right\vert \approx 0.03$~s$^{-1}$, and, if
$\gamma _{L}/2\pi \approx 100$~Hz, at $\left\vert \Omega _{0}\right\vert >0.3
$~s$^{-1}$. The number of photons emitted at the absorption saturation reaches
the maximum value of $\gamma /2\sim 0.0001$~s$^{-1}$ per each nucleus on
the average.

Since the nucleus spin changes from 5/2 to 3/2 at the excitation of the nucleus, the
hyperfine structure of energy levels of an atom or ion with the nucleus in
the isomeric state also changes. As a result, there emerges a possibility of
detecting the excited nucleus with the help of optical laser radiation with
the frequency tuned to one of the transition frequencies between hyperfine
levels. Since the linewidths $\Gamma $'s for the corresponding electron
transitions are of the order of 10$^{7}~\mathrm{s}^{-1}$, the excitation of
each nucleus stimulates the emission of about 10$^{7}$~photons during a
second. The implementation of this registration setup requires that atoms or
ions should be excited simultaneously from all sublevels of the hyperfine
structure of the ground electron state of an atom or ion with the isomeric
nucleus in order to prohibit atoms or ions to be accumulated on one of them
when the transition from the excited state into the ground one takes place
in the course of spontaneous radiation emission.

Let us evaluate the Rabi frequency at the two-photon excitation of
$\mathrm{Th}^{3+}$ ion. Its ground state has six hyperfine components with the total
moment $F$ ranging from 0 to 5, and the ground state of the electron
shell of an ion with the excited nucleus has four hyperfine components with the
$F$-values ranging from 1 to 4. From expressions~(\ref{eq:Rabi-shift-ii-I}),
(\ref{eq:Omega0-ii-I}), and (\ref{eq:Omega0av}), one can see that the Rabi
frequency is maximal, if the product $(\mu _{ee}-\mu _{gg})\mu _{eg}$ is
maximal. This condition is satisfied for the transition between the levels
$|g\rangle =|F=1,m=0,I=5/2\rangle $ and $|e\rangle =|F=1,m=1,I=3/2\rangle $.
In this case,
%65
\begin{equation}
\mu _{ee}-\mu _{gg}=\frac{3}{2}\mu _{\rm B},
\end{equation}
%66
\[
 \mu_{eg}=(-1)^{F_g+I_e+J-1}\|\mu_{eg}\|\sqrt{2F_{g}+1}\times
\]
\begin{equation}
\times C_{F_gm_g1q}^{F_em_e} \left\{
\begin{array}{ccc}
I_g & J& F_g \\
F_e & 1 & I_e
\end{array}
 \right\}=-\sqrt{\frac{7}{30}}\frac{\|\mu_{eg}\|}{2}.
\end{equation}
For a laser radiation power of 100~mW and provided that radiation is focused
into a spot 1~$\mu \mathrm{m}$ in diameter, we obtain
$I=10^{7}$~\textrm{W/}$\mathrm{cm}^{\mathrm{2}}$, with the Rabi frequency of the transition, according
to Eq.~(\ref{eq:Omega0av}), being equal to $\Omega _{0}\approx
0.07~\mathrm{s}^{-1}$ and the light shift, according to Eq.~(\ref{eq:S0av}), being equal
to about $5.3\times 10^{-6}~\mathrm{s}^{-1}$. So a small value of light
shift testifies that the very procedure of recording the signal from a
thorium nucleus inserts, in essence, no additional error into the
results of transition frequency measurements. The actual accuracy of optical
clocks is governed by other factors, which are not associated with the thorium
excitation; in particular, these are the frequency stability of a radiation
source and the field in the trap. The indicated evaluation of the Rabi frequency
testifies to a capability of achieving the absorption saturation at the two-photon
transition in a thorium nucleus by applying moderate-power laser radiation.

In order to enhance the fluorescence signal, it is necessary to increase the
number of irradiated thorium nuclei. A high concentration of thorium nuclei
can be achieved in crystals. In this case, on the one hand, the Rabi
frequency becomes lower by three orders of magnitude, because thorium is
tetravalent in known chemical compounds. On the other hand, however, the
number of thorium ions can be increased by many orders of magnitude.
Let us evaluate the number of emitted photons in this case. In the case of
the transition in $\mathrm{Th}^{4+}$, the expression for the transition dipole
moment looks like
%67
\begin{equation}
\mu _{eg}=\frac{\|\mu _{eg}\|}{\sqrt{2I_{e}+1}}
C_{I_{g}m_{g}1q}^{I_{e}m_{e}},\quad q=m_{e}-m_{g}.
\end{equation}%
For $|g\rangle =|I=5/2,m=3/2\rangle $ and $|e\rangle =|I=3/2,m=3/2\rangle $,
we obtain $\mu _{eg}=\frac{1}{\sqrt{15}}\left\Vert \mu _{eg}\right\Vert $.
For a laser radiation power of 100~mW and if radiation is focused
into a spot 1~$\mu \mathrm{m}$ in diameter, we obtain
$I=10^{7}$~\textrm{W/}$\mathrm{cm}^{\mathrm{2}}$. Such a radiation intensity cannot guarantee
the absorption saturation. The Rabi frequency $\Omega _{0}$ of the transition,
according to Eq.~(\ref{eq:Omega0av}), is of the order of $1\times
10^{-5}~\mathrm{s}^{-1}$, and the term containing $\Omega _{0}$ in the
denominator of expression~(\ref{eq:sat}) for $\rho _{22}$ can be neglected.

Let the crystal be illuminated with a laser beam characterized by the
spatial intensity distribution
%68
\begin{equation}
I=\frac{2P}{\pi {}w^{2}}\exp \left( -\frac{2r^{2}}{w^{2}}\right),
\label{eq:G-I}
\end{equation}
where $P$ is the laser radiation power,
%69
\begin{equation}
w=w_{0}\sqrt{1+\frac{z^{2}}{b^{2}}},  \label{eq:w}
\end{equation}%
$r$ is the distance from the point to the beam axis, $w_{0}$ the minimum
beam radius (the minimum distance from the beam axis, at which the field
decreases by a factor of $e$), $z$ the coordinate along the beam, $b=\pi
w_{0}^{2}/\lambda $ is the confocal parameter, and $\lambda $ the wavelength
of laser radiation. Let us determine the number of photons emitted per unit
time from the irradiated crystal volume at the two-photon excitation. Let the
concentration of thorium nuclei in the specimen be $n$. Then $\gamma n\rho
_{22}$ photons are emitted per unit volume per unit time. Taking the
inequality $\Omega _{0}\ll \gamma \gamma ^{\prime }$ into account and
integrating over the whole specimen volume, we obtain the total number of
emitted photons per unit time,
%70
\begin{equation}
F=\int\limits_{0}^{\infty }dr\,2\pi {}r\int\limits_{-\infty }^{\infty
}dz\,{n\mathcal{N}}=\frac{n\pi {}P^{2}}{2\lambda \gamma ^{\prime }}\left[ \frac{\mu
_{ge}(\mu _{ee}-\mu _{gg})}{\hbar ^{2}\omega {}c}\mu _{0}\right] ^{2}.
\label{eq:FlF}
\end{equation}%
In work \cite{Tka11}, while estimating the possibility of the creation of
an optical $\gamma $-laser on a $\mathrm{LiCaAlF_{6}}$ crystal with a thorium
impurity, the concentration of the latter was taken to equal
$10^{18}$~\textrm{cm}$^{-3}$, so that Th$^{4+}$ ions did not change the crystal
structure substantially. For our estimation, we adopt, as was done in the
case of free ion, that $\gamma ^{\prime }=\gamma _{L}$. Then, according to
Eq.~(\ref{eq:FlF}), the specimen emits 3.6 photons per second if the
radiation intensity is 10~W and $\gamma _{L}/2\pi =1$~Hz.

Now, let us evaluate the volume of the active region in the crystal. For
$w_{0}=10$~$\mu \mathrm{m}$ and $\lambda =320$\textrm{~nm,} we
obtain $b=1$~mm. Therefore, the crystal volume can be smaller than
0.1~mm$^{3}$. Provided that the concentration equals
$n=10^{18}$~\textrm{cm}$^{-3}$, such a volume contains
$n=10^{14}$~thorium atoms, 400 of which decay every second. As a
result, additional radiation may be generated, by depending on the
crystal used. For example, in a $\mathrm{CaF_{2}}$ crystal
25~mm$^{3}$ in dimension, every $\alpha $-decay of
$^{241}\mathrm{Am}$ invoked scintillations (about 40~photons) during
10$^{-5}~\mathrm{s}$ in the spectral range of 220--400\textrm{~nm}
far from the length of the excited $^{229}\mathrm{Th}$ fluorescence
signal~\cite{Mik06}. This radiation can be reduced considerably by
applying a thin specimen in order to allow $\alpha $-particles to
leave it quickly after the decay. In particular, a crystal 1~$\mu
\mathrm{m}$ in thickness on a metal substrate can be used, in which
the fluorescence can be excited with the help of a surface
electromagnetic wave.

\section{Conclusions}

Our analysis of two-photon optical transitions in a
$^{229}\mathrm{Th}$ nucleus showed that the averaged intensities of
a propagating monochromatic wave and the field of a sequence of
short light pulses, which are required for the identical excitation
of nuclei, are equal to each other. The light shifts in the
polychromatic and monochromatic fields are also identical. Since
radiation for the excitation of nuclei (with a wavelength of about
320\textrm{~nm}) is generated in multiphoton processes, it is easier
to obtain the required intensity of laser radiation from a pulse
sequence rather than from monochromatic radiation. Moreover, the
application of two-photon transitions allows the background
scattering signal to be reduced substantially, because nuclei are
excited at a frequency that is half as high as that for the
fluorescence signal. Another advantage of the pulse excitation
consists in a possibility of measuring the frequency of the nuclear
transition, provided that the excitation is carried out with the
help of a broadband frequency comb. While irradiating Th$^{3+}$
thorium ions in an electromagnetic trap and at a pumping power of
100~mW, the absorption saturation at the two-photon transition can
be attained. With regard for the long lifetime of excited nuclei at
their detection, an auxiliary radiation can be used, which is in
resonance with one of the transitions between the components of the
hyperfine structure of a thorium ion with the isomeric nucleus. This
circumstance allows the fluorescence signal to be enhanced by
several orders of magnitude. If a 10-W laser radiation is used to
excite Th$^{4+}$ ions in a solid, several photons per second can
expectedly be emitted with a frequency equal to that of the optical
transition in a thorium-229 nucleus. Thus, the method proposed here
for the creation of excited isomeric state of thorium-229 on the
basis of the two-photon absorption can be used in both solid-state
optical nuclear clocks and clocks on
the basis of ions in an electromagnetic trap.%\looseness=1

\vskip3mm

This work was supported by the State Fund for Fundamental Researches of
Ukraine (project~F40.2/039), the Austrian Science Fund (FWF,
project~M1272-N16), the Federal Goal-oriented Program \textquotedblleft
Scientific and pedagogical staff for innovative Russia in
2009--2013\textquotedblright\ (grants GK~14.740.11.0463 and
GK~16.740.11.0586), and the Russian Foundation for Basic Research (grant
RFFI-11-02-90426\_Ukr\_F\_a).

 \rezume{ПРЯМЕ ДВОФОТОННЕ ЗБУДЖЕННЯ  ІЗОМЕРНОГО\\
ПЕРЕХОДУ
В ЯДРІ ТОРІЮ-229}{В.І. Романенко, О.Г. Удовицька, Л.П. Яценко,\\
О.В. Романенко, А.Н. Літвінов, Г.А. Казаков}{ Розглядається
можливість двофотонного збудження ізомерного стану в ядрі торію-229.
Показано, що інтенсивність флуоресценції однакова при збудженні ядер
монохроматичним випромінюванням або поліхроматичним випромінюванням
послідовності коротких світлових імпульсів тієї ж інтенсивності. При
двофотонному збудженні іона $ \mathrm{Th^{3 +}} $ в електромагнітній
пастці сфокусованим випромінюванням лазера з довжиною хвилі $
\sim320 $~нм і потужністю близько 100~мВт можна досягти насичення
поглинання, за якого випромінювання флуоресценції з частотою
переходу в ядрі максимальне. В кристалах, допованих $
\mathrm{Th^{4+}} $ з концентрацією близько $ 10^{18} $~см$^{-3} $, у
полі лазерного випромінювання потужністю 10 Вт можливе
випромінювання кількох фотонів за секунду з довжиною хвилі $ \sim160
$~нм. }

\end{document}